\documentclass{article}

\usepackage[final]{nips_2018}

\usepackage[utf8]{inputenc} 
\usepackage[T1]{fontenc}    
\usepackage{url}            
\usepackage{booktabs}       
\usepackage{amsfonts}       
\usepackage{nicefrac}       
\usepackage{microtype}      

\usepackage{amsmath}
\usepackage{amssymb}
\usepackage{floatrow}

\usepackage{color}
\usepackage{graphicx}

\usepackage{bm}

\newcommand{\pn}{\mathrm{P}}
\newcommand{\intn}{\mathrm{I}}
\newcommand{\leak}{\mathrm{lk}}
\newcommand{\dnd}{\mathrm{D}}
\newcommand{\apic}{\mathrm{A}}
\newcommand{\bas}{\mathrm{B}}
\newcommand{\som}{\mathrm{som}}
\newcommand{\exc}{\mathrm{exc}}
\newcommand{\inh}{\mathrm{inh}}
\newcommand{\rest}{\mathrm{rest}}
\newcommand{\topdown}{\mathrm{TD}}

\DeclareMathOperator{\Tr}{Tr}
\DeclareMathOperator{\Expect}{E}

\title{Dendritic cortical microcircuits\\ approximate the backpropagation algorithm}

\author{
  João Sacramento\thanks{Present address: Institute of Neuroinformatics, University of Zürich and ETH Zürich, Zürich, Switzerland}\\
  Department of Physiology\\
  University of Bern, Switzerland\\
  \texttt{sacramento@pyl.unibe.ch} \\
  \And
  Rui Ponte Costa\thanks{Present address: Computational Neuroscience Unit, Department of Computer Science, SCEEM, Faculty of Engineering, University of Bristol, United Kingdom}\\
  Department of Physiology\\
  University of Bern, Switzerland\\
  \texttt{costa@pyl.unibe.ch}\\
  \And
  Yoshua Bengio\thanks{CIFAR Senior Fellow}\\
  Mila and Université de Montréal, Canada\\
  \texttt{yoshua.bengio@mila.quebec} \\
  \And
  Walter Senn\\
  Department of Physiology\\
  University of Bern, Switzerland\\
  \texttt{senn@pyl.unibe.ch}\\
}

\begin{document}

\maketitle

\begin{abstract}
Deep learning has seen remarkable developments over the last years, many of them inspired by neuroscience. However, the main learning mechanism behind these advances -- error backpropagation -- appears to be at odds with neurobiology.
Here, we introduce a multilayer neuronal network model with simplified dendritic compartments in which error-driven synaptic plasticity adapts the network towards a global desired output. In contrast to previous work our model does not require separate phases and synaptic learning is driven by local dendritic prediction errors continuously in time. Such errors originate at apical dendrites and occur due to a mismatch between predictive input from lateral interneurons and activity from actual top-down feedback. Through the use of simple dendritic compartments and different cell-types our model can represent both error and normal activity within a pyramidal neuron. We demonstrate the learning capabilities of the model in regression and classification tasks, and show analytically that it approximates the error backpropagation algorithm. Moreover, our framework is consistent with recent observations of learning between brain areas and the architecture of cortical microcircuits. Overall, we introduce a novel view of learning on dendritic cortical circuits and on how the brain may solve the long-standing synaptic credit assignment problem.
\end{abstract}

\section{Introduction}

Machine learning is going through remarkable developments powered by deep neural networks  \citep{LeCun2015}. Interestingly, the workhorse of deep learning is still the classical backpropagation of errors algorithm \citep[backprop;][]{Rumelhart1986}, which has been long dismissed in neuroscience on the grounds of biologically implausibility \citep{Grossberg1987,Crick1989}. Irrespective of such concerns, growing evidence demonstrates that deep neural networks outperform alternative frameworks in accurately reproducing activity patterns observed in the cortex \citep{Lillicrap2013,Yamins2014,Ravazi2014,Yamins2016,Kell2018}. Although recent developments have started to bridge the gap between neuroscience and artificial intelligence \citep{Marblestone:2016bm,Lillicrap2016,Scellier2017,costa2017cortical,Guerguiev2017}, how the brain could implement a backprop-like algorithm remains an open question.

In neuroscience, understanding how the brain learns to associate different areas (e.g., visual and motor cortices) to successfully drive behaviour is of fundamental importance \citep{Petreanu:2012dy,Manita2015,Makino:2015jr,Poort:2015fwa,Fu:2015hn,Pakan:2016bm,Zmarz:2016jo,Attinger:2017bo}. However, how to correctly modify synapses to achieve this has puzzled neuroscientists for decades. This is often referred to as the synaptic credit assignment problem \citep{Rumelhart1986,sutton1998reinforcement,Roelfsema2005,Friedrich:2011jm,Bengio2014,Lee2015,Roelfsema2018}, for which the backprop algorithm provides an elegant solution.

Here we propose that the prediction errors that drive learning in backprop are encoded at distal dendrites of pyramidal neurons, which receive top-down input from downstream brain areas (we interpret a brain area as being equivalent to a layer in machine learning) \citep{Petreanu:2009kka,Larkum2013}. In our model, these errors arise from the inability to exactly match via lateral input from local interneurons (e.g.~somatostatin-expressing; SST) the top-down feedback from downstream cortical areas. Learning of bottom-up connections (i.e., feedforward weights) is driven by such error signals through local synaptic plasticity. Therefore, in contrast to previous approaches \citep{Marblestone:2016bm}, in our framework a given neuron is used simultaneously for activity propagation (at the somatic level), error encoding (at distal dendrites) and error propagation to the soma without the need for separate phases.

We first illustrate the different components of the model. Then, we show analytically that under certain conditions learning in our network approximates backpropagation. Finally, we empirically evaluate the performance of the model on nonlinear regression and recognition tasks.

\vspace{-2mm}
\section{Error-encoding dendritic cortical microcircuits}

\subsection{Neuron and network model}

\begin{figure}[htb!]
  \centering
  \includegraphics[width=1\linewidth]{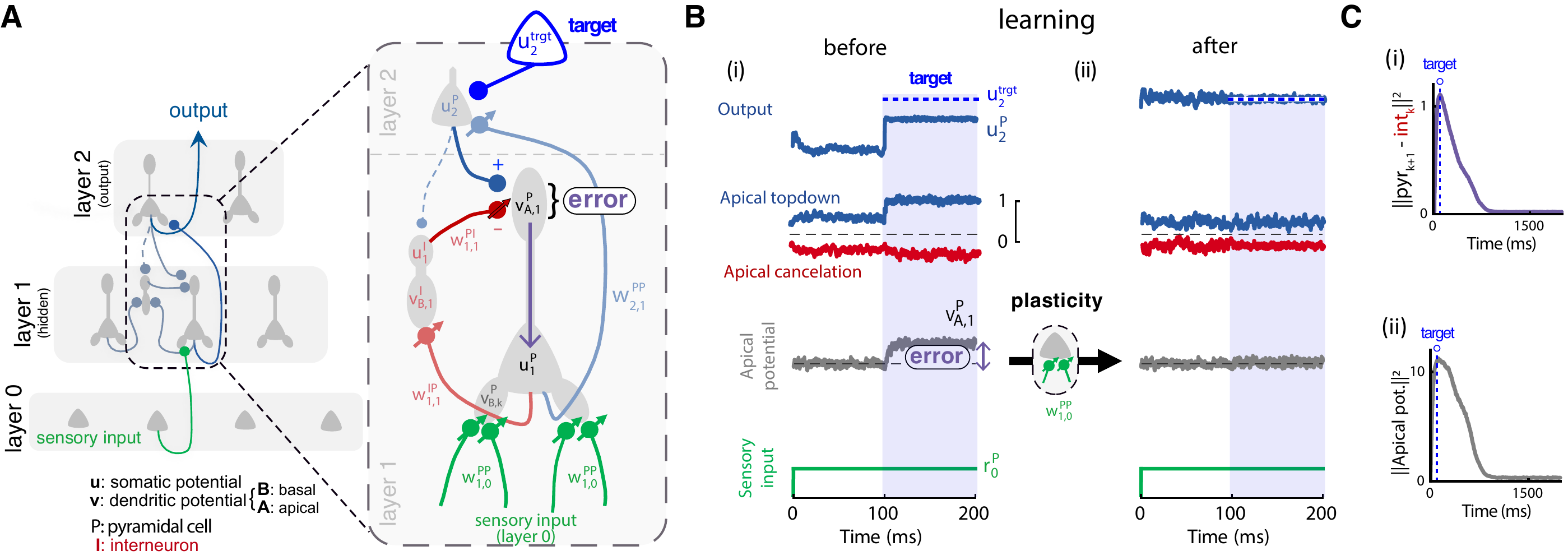}
  \caption{\textbf{Learning in error-encoding dendritic microcircuit network.}
  (\textbf{A}) Schematic of network with pyramidal cells and lateral inhibitory interneurons. Starting from a self-predicting state -- see main text and supplementary material (SM) -- when a novel teaching (or associative) signal is presented at the output layer ($\mathbf{u}_2^{\mathrm{trgt}}$), a prediction error in the apical compartments of pyramidal neurons in the upstream layer (layer 1, `error') is generated. This error appears as an apical voltage deflection that propagates down to the soma (purple arrow) where it modulates the somatic firing rate, which in turn leads to plasticity at bottom-up synapses (bottom, green).
  (\textbf{B}) Activity traces in the microcircuit before and after a new teaching signal is learned. (i) Before learning: a new teaching signal is presented ($\mathbf{u}_2^\mathrm{trgt}$), which triggers a mismatch between the top-down feedback (grey blue) and the cancellation given by the lateral interneurons (red). (ii) After learning (with plasticity at the bottom-up synapses ($\mathbf{W}_{1,0}^{\pn\pn}$)), the network successfully predicts the new teaching signal, reflected on no distal 'error' (top-down and lateral interneuron input cancel each other).
  (\textbf{C}) Interneurons learn to predict the backpropagated activity (i), while simultaneously silencing the apical compartment (ii), even though the pyramidal neurons remain active (not shown).}
  \label{fig:learning_newinput}
\end{figure}

Building upon previous work \citep{Urbanczik2014}, we adopt a simplified multicompartment neuron and describe pyramidal neurons as three-compartment units (schematically depicted in Fig.~\ref{fig:learning_newinput}A). These compartments represent the somatic, basal and apical integration zones that characteristically define neocortical pyramidal cells \citep{Spruston2008,Larkum2013}. The dendritic structure of the model is exploited by having bottom-up and top-down synapses converging onto separate dendritic compartments (basal and distal dendrites, respectively), a first approximation in line with experimental observations \citep{Spruston2008} and reflecting the preferred connectivity patterns of cortico-cortical projections \citep{Larkum2013}.

Consistent with the connectivity of SST interneurons \citep{UrbanCiecko:2016io}, we also introduce a second population of cells within each hidden layer with both lateral and cross-layer connectivity, whose role is to {\em cancel the top-down input} so as to leave only the backpropagated errors as apical dendrite activity. Modelled as two-compartment units (depicted in red, Fig.~\ref{fig:learning_newinput}A), such interneurons are predominantly driven by pyramidal cells within the same layer through weights $\mathbf{W}_{k,k}^{\intn\pn}$, and they project back to the apical dendrites of the same-layer pyramidal cells through weights $\mathbf{W}_{k,k}^{\pn\intn}$ (Fig.~\ref{fig:learning_newinput}A). Additionally, cross-layer feedback onto SST cells originating at the next upper layer $k\!+\!1$ provide a weak nudging signal for these interneurons, modelled after \citet{Urbanczik2014} as a conductance-based somatic input current. We modelled this weak top-down nudging on a one-to-one basis: each interneuron is nudged towards the potential of a corresponding upper-layer pyramidal cell. Although the one-to-one connectivity imposes a restriction in the model architecture, this is to a certain degree in accordance with recent monosynaptic input mapping experiments show that SST cells in fact receive top-down projections \citep{Leinweber:2017jq}, that according to our proposal may encode the weak interneuron `teaching' signals from higher to lower brain areas.

The somatic membrane potentials of pyramidal neurons and interneurons evolve in time according to
\begin{align}
  \label{eq:dUPdt} \frac{d}{d t}{\mathbf{u}}_k^\pn(t) &= - g_\leak \, \mathbf{u}_k^\pn(t) + g_\bas \left(\mathbf{v}_{\bas, k}^\pn(t) - \mathbf{u}_k^\pn(t) \right) + g_\apic \left(\mathbf{v}_{\apic, k}^\pn(t) - \mathbf{u}_k^\pn(t)\right) + \sigma \, \boldsymbol{\xi}(t)\\
  \label{eq:dUIdt} \frac{d}{d t}{\mathbf{u}}_k^\intn(t)  &= - g_\leak \, \mathbf{u}_k^\intn(t) + g_\dnd \left(\mathbf{v}_{k}^\intn(t) - \mathbf{u}_k^\intn(t) \right) + \mathbf{i}^\intn_k(t) + \sigma \, \boldsymbol{\xi}(t),
\end{align}
with one such pair of dynamical equations for every hidden layer $0 < k < N$; input layer neurons are indexed by $k = 0$, $g$'s are fixed conductances, $\sigma$ controls the amount of injected noise. Basal and apical dendritic compartments of pyramidal cells are coupled to the soma with effective transfer conductances $g_\bas$ and $g_\apic$, respectively. Subscript $\leak$ is for leak, $\apic$ is for apical, $\bas$ for basal, $\dnd$ for dendritic, superscript $\intn$ for inhibitory and $\pn$ for pyramidal neuron. Eqs.~\ref{eq:dUPdt} and \ref{eq:dUIdt} describe standard conductance-based voltage integration dynamics, having set membrane capacitance to unity and resting potential to zero for clarity. Background activity is modelled as a Gaussian white noise input, $\boldsymbol{\xi}$ in the equations above. To keep the exposition brief we use matrix notation, and denote by $\mathbf{u}_k^\pn$ and $\mathbf{u}_k^\intn$ the vectors of pyramidal and interneuron somatic voltages, respectively. Both matrices and vectors, assumed column vectors by default, are typed in boldface here and throughout. Dendritic compartmental potentials are denoted by $\mathbf{v}$ and are given in instantaneous form by
\begin{align}
  \label{eq:Vbas} \mathbf{v}_{\bas, k}^\pn(t) &= \mathbf{W}^{\pn \pn}_{k,k-1} \, \phi(\mathbf{u}_{k-1}^\pn(t)) \\
  \label{eq:Vapic} \mathbf{v}_{\apic, k}^\pn(t) &= \mathbf{W}^{\pn \pn}_{k,k+1} \, \phi(\mathbf{u}_{k+1}^\pn(t)) + \mathbf{W}^{\pn \intn}_{k,k} \, \phi(\mathbf{u}_{k}^\intn(t)),
\end{align}
where $\phi(\mathbf{u})$ is the neuronal transfer function, which acts componentwise on $\mathbf{u}$.

For simplicity, we reduce pyramidal output neurons to two-compartment cells: the apical compartment is absent ($g_\apic = 0$ in Eq.~\ref{eq:dUPdt}) and basal voltages are as defined in Eq.~\ref{eq:Vbas}. Although the design can be extended to more complex morphologies, in the framework of dendritic predictive plasticity two compartments suffice to compare desired target with actual prediction. Synapses proximal to the soma of output neurons provide direct external teaching input, incorporated as an additional source of current $\mathbf{i}^\pn_N$. In practice, one can simply set $\mathbf{i}^\pn_N = g_\som (\mathbf{u}_N^\text{trgt} - \mathbf{u}_N^\pn)$, with some fixed somatic nudging conductance $g_\som$. This can be modelled closer to biology by explicitly setting the somatic excitatory and inhibitory conductance-based inputs \citep{Urbanczik2014}. For a given output neuron, $i_N^\pn(t) = g_{\exc,N}^\pn(t) \left(E_\exc - u_N^\pn(t)\right) + g_{\inh,N}^\pn(t) \left(E_\inh - u_N^\pn(t)\right)$, where $E_\exc$ and $E_\inh$ are excitatory and inhibitory synaptic reversal potentials, respectively, where the inputs are balanced according to $g_{\exc, N}^\pn = g_\som \frac{u_{N}^\text{trgt} - E_\inh}{E_\exc - E_\inh}$, $g_{\exc, N}^\pn = - g_\som \frac{u_{N}^\text{trgt} - E_\exc}{E_\exc - E_\inh}$. The point at which no current flows, $i^\pn_N = 0$, defines the target teaching voltage $u_N^\text{trgt}$ towards which the neuron is nudged\footnote{Note that in biology a target may be represented by an associative signal from the motor cortex to a sensory cortex \citep{Attinger:2017bo}.}.

Interneurons are similarly modelled as two-compartment cells, cf.~Eq.~\ref{eq:dUIdt}. Lateral dendritic projections from neighboring pyramidal neurons provide the main source of input as
\begin{equation}
  \label{eq:Vbasintn} \mathbf{v}_{k}^\intn(t) = \mathbf{W}^{\intn \pn}_{k,k} \, \phi(\mathbf{u}_{k}^\pn(t)),
\end{equation}
whereas cross-layer, top-down synapses define the teaching current $\mathbf{i}^\intn_k$. This means that an interneuron at layer $k$ permanently (i.e., when learning or performing a task) receives balanced somatic teaching excitatory and inhibitory input from a pyramidal neuron at layer $k\!+\!1$ on a one-to-one basis (as above, but with $\mathbf{u}^\pn_{k+1}$ as target). With this setting, the interneuron is nudged to follow the corresponding next layer pyramidal neuron. See SM for detailed parameters.

\vspace{-0.5mm}
\subsection{Synaptic learning rules}
The synaptic learning rules we use belong to the class of dendritic predictive plasticity rules \citep{Urbanczik2014,Spicher2017} that can be expressed in its general form as
\begin{equation}
  \label{eq:dwdt}
  \frac{d}{dt} w = \eta \, \left(\phi(u) - \phi(v) \right) r,
\end{equation}
where $w$ is an individual synaptic weight, $\eta$ is a learning rate, $u$ and $v$ denote distinct compartmental potentials, $\phi$ is a rate function, and $r$ is the presynaptic input. Eq.~\ref{eq:dwdt} was originally derived in the light of reducing the prediction error of somatic spiking, when $u$ represents the somatic potential and $v$ is a function of the postsynaptic dendritic potential.

In our model the plasticity rules for the various connection types are:
\begin{align}
  \label{eq:dW-PP}\frac{d}{dt} \mathbf{W}^{\pn\pn}_{k,k-1} &= \eta^{\pn\pn}_{k,k-1} \left(\phi(\mathbf{u}^\pn_k) - \phi(\hat{\mathbf{v}}^{\pn}_{\bas,k}) \right) \left(\mathbf{r}_{k-1}^\pn\right)^{\! T},\\
  \label{eq:dW-IP}\frac{d}{dt} \mathbf{W}^{\intn\pn}_{k,k} &= \eta^{\intn\pn}_{k,k} \left(\phi(\mathbf{u}^\intn_k) - \phi(\hat{\mathbf{v}}^{\intn}_{k}) \right) \left(\mathbf{r}_{k}^\pn\right)^{\! T},\\
  \label{eq:dW-PI}\frac{d}{dt} \mathbf{W}^{\pn\intn}_{k,k} &= \eta^{\pn\intn}_{k,k} \left(\mathbf{v}_\rest - \mathbf{v}^{\pn}_{\apic,k} \right) \left(\mathbf{r}_{k}^\intn\right)^{\! T},
\end{align}
where $(\cdot)^T$ denotes vector transpose and $\mathbf{r}_k \equiv \phi(\mathbf{u}_k)$ the layer $k$ firing rates. The synaptic weights evolve according to the product of dendritic prediction error and presynaptic rate, and can undergo both potentiation or depression depending on the sign of the first factor (i.e., the prediction error).

For basal synapses, such prediction error factor amounts to a difference between postsynaptic rate and a local dendritic estimate which depends on the branch potential. In Eqs.~\ref{eq:dW-PP} and ~\ref{eq:dW-IP},  $\hat{\mathbf{v}}^{\pn}_{\bas,k} = \frac{g_\bas}{g_\leak + g_\bas + g_\apic} \, \mathbf{v}^{\pn}_{\bas,k}$ and $\hat{\mathbf{v}}^{\intn}_{k} = \frac{g_\dnd}{g_\leak + g_\dnd} \, \mathbf{v}^{\intn}_{k}$ take into account dendritic attenuation factors of the different compartments. On the other hand, the plasticity rule \eqref{eq:dW-PI} of lateral interneuron-to-pyramidal synapses aims to silence (i.e., set to resting potential $\mathbf{v}_\rest = \mathbf{0}$, here and throughout zero for simplicity) the apical compartment; this introduces an attractive state for learning where the contribution from interneurons balances (or cancels out) top-down dendritic input. This learning rule of apical-targeting interneuron synapses can be thought of as a dendritic variant of the homeostatic inhibitory plasticity proposed by \citet{Vogels2011,Luz2012}.

In experiments where the top-down connections are plastic, the weights evolve according to
\begin{equation}
  \label{eq:dW-PP-TD}\frac{d}{dt} \mathbf{W}^{\pn\pn}_{k,k+1} = \eta^{\pn\pn}_{k,k+1} \left(\phi(\mathbf{u}^\pn_k) - \phi(\hat{\mathbf{v}}_{\topdown,k}^\pn) \right) \left(\mathbf{r}_{k+1}^\pn\right)^{\! T},
\end{equation}
with $\hat{\mathbf{v}}_{\topdown,k}^\pn = \mathbf{W}_{k,k+1} \, \mathbf{r}_{k+1}^{\pn}$.
An implementation of this rule requires a subdivision of the apical compartment into a distal part receiving the top-down input (with voltage $\hat{\mathbf{v}}_{\topdown,k}^\pn$) and another distal compartment receiving the lateral input from the interneurons (with voltage $\mathbf{v}_{\apic,k}^\pn)$.

\vspace{-0.1cm}
\subsection{Comparison to previous work}
It has been suggested that error backpropagation could be approximated by an algorithm that requires alternating between two learning phases, known as contrastive Hebbian learning \citep{Ackley1985}. This link between the two algorithms was first established for an unsupervised learning task \citep{Hinton1988} and later analyzed \citep{Xie2003} and generalized to broader classes of models \citep{OReilly1996,Scellier2017}.

The concept of apical dendrites as distinct integration zones, and the suggestion that this could simplify the implementation of backprop has been previously made \citep{Kording2000,Kording2001}. Our microcircuit design builds upon this view, offering a concrete mechanism that enables apical error encoding. In a similar spirit, two-phase learning recently reappeared in a study that exploits dendrites for deep learning with biological neurons \citep{Guerguiev2017}.
 In this more recent work, the temporal difference between the activity of the apical dendrite in the presence and in the absence of the teaching input represents the error that induces plasticity at the forward synapses. This difference is used directly for learning the bottom-up synapses without influencing the somatic activity of the pyramidal cell. In contrast, we postulate that the apical dendrite has an explicit error representation by simultaneously integrating top-down excitation and lateral inhibition. As a consequence, we do not need to postulate separate temporal phases, and our network operates continuously while plasticity at all synapses is always turned on.

Error minimization is an integral part of brain function according to predictive coding theories \citep{Rao1999,Friston2005}. Interestingly, recent work has shown that backprop can be mapped onto a predictive coding network architecture \citep{Whittington:2017js}, related to the general framework introduced by \citet{LeCun1988}. A possible network implementation is suggested by \citet{Whittington:2017js} that requires intricate circuitry with appropriately tuned error-representing neurons. According to this work, the only plastic synapses are those that connect prediction and error neurons. By contrast, in our model, lateral, bottom-up and top-down connections are all plastic, and errors are directly encoded in dendritic compartments.

\section{Results}

\subsection{Learning in dendritic error networks approximates backprop}

In our model, neurons implicitly carry and transmit errors across the network. In the supplementary material, we formally show such propagation of errors for networks in a particular regime, which we term \emph{self-predicting}. Self-predicting nets are such that when no external target is provided to output layer neurons, the lateral input from interneurons cancels the internally generated top-down feedback and renders apical dendrites silent. In this case, the output becomes a feedforward function of the input, which can in theory be optimized by conventional backprop. We demonstrate that synaptic plasticity in self-predicting nets approximates the weight changes prescribed by backprop.

We summarize below the main points of the full analysis (see SM). First, we show that somatic membrane potentials at hidden layer $k$ integrate feedforward predictions (encoded in basal dendritic potentials) with backpropagated errors (encoded in apical dendritic potentials):
\begin{align}
  \mathbf{u}^{\pn}_k &= \mathbf{u}_k^- + \lambda^{N-k+1} \, \mathbf{W}_{k,k+1}^{\pn\pn} \, \left(\prod_{l=k+1}^{N-1} \mathbf{D}_{l}^- \, \mathbf{W}_{l, l+1}^{\pn\pn} \right) \mathbf{D}_{N}^- \left(\mathbf{u}_N^\text{trgt} - \mathbf{u}_N^- \right) + \mathcal{O}(\lambda^{N-k+2})\nonumber.
\end{align}
Parameter $\lambda \ll 1$ sets the strength of feedback and teaching versus bottom-up inputs and is assumed to be small to simplify the analysis. The first term is the basal contribution and corresponds to $\mathbf{u}_k^-$, the activation computed by a purely feedforward network that is obtained by removing lateral and top-down weights from the model (here and below, we use superscript `-' to refer to the feedforward model). The second term (of order $\lambda^{N-k+1}$) is an error that is backpropagated from the output layer down to $k$-th layer hidden neurons; matrix $\mathbf{D}_k$ is a diagonal matrix with $i$-th entry containing the derivative of the neuronal transfer function evaluated at $u_{k,i}^-$.

Second, we compare model synaptic weight updates for the bottom-up connections to those prescribed by backprop. Output layer updates are exactly equal by construction. For hidden neuron synapses, we obtain
\begin{align}
  \Delta \mathbf{W}_{k,k-1}^{\pn\pn} &= \eta_{k,k-1}^{\pn\pn} \lambda^{N-k+1} \left(\prod_{l=k}^{N-1} \mathbf{D}_{l}^- \, \mathbf{W}_{l, l+1}^{\pn\pn} \right) \mathbf{D}_{N}^- \left(\mathbf{u}_N^\text{trgt} - \mathbf{u}_N^- \right) \left(\mathbf{r}_{k-1}^-\right)^T + \mathcal{O}(\lambda^{N-k+2}).\nonumber
\end{align}
Up to a factor which can be absorbed in the learning rate, this plasticity rule becomes equal to the backprop weight change in the weak feedback limit $\lambda \to 0$, provided that the top-down weights are set to the transpose of the corresponding feedforward weights.

In our simulations, top-down weights are either set at random and kept fixed, in which case the equation above shows that the plasticity model optimizes the predictions according to an approximation of backprop known as feedback alignment \citep{Lillicrap2016}; or learned so as to minimize an inverse reconstruction loss, in which case the network implements a form of target propagation \citep{Bengio2014,Lee2015}.

\subsection{Deviations from self-predictions encode backpropagated errors}

To illustrate learning in the model and to confirm our analytical insights we first study a very simple task: memorizing a single input-output pattern association with only one hidden layer; the task naturally generalizes to multiple memories.

Given a self-predicting network (established by microcircuit plasticity, Fig.~\ref{fig:learn-self-predicting-state}, see SM for more details), we focus on how prediction errors get propagated backwards when a novel teaching signal is provided to the output layer, modeled via the activation of additional somatic conductances in output pyramidal neurons. Here we consider a network model with an input, a hidden and an output layer (layers 0, 1 and 2, respectively; Fig.~\ref{fig:learning_newinput}A).

When the pyramidal cell activity in the output layer is nudged towards some desired target (Fig.~\ref{fig:learning_newinput}B (i)), the bottom-up synapses $\mathbf{W}_{2,1}^{\pn\pn}$ from the lower layer neurons to the basal dendrites are adapted, again according to the plasticity rule that implements the dendritic prediction of somatic spiking (see Eq.~\ref{eq:dW-PP}). What these synapses cannot explain away encodes a dendritic error in the pyramidal neurons of the lower layer 1. In fact, the self-predicting microcircuit can only cancel the feedback that is produced by the lower layer activity.

The somatic integration of apical activity induces plasticity at the bottom-up synapses $\mathbf{W}_{1,0}^{\pn\pn}$ (Eq.~\ref{eq:dW-PP}). As the apical error changes the somatic activity, plasticity of the $\mathbf{W}_{1,0}^{\pn\pn}$ weights tries to further reduce the error in the output layer. Importantly, the plasticity rule depends only on local information available at the synaptic level: postsynaptic firing and dendritic branch voltage, as well as the presynaptic activity, in par with phenomenological models of synaptic plasticity \citep{Sjostrom2001,Clopath2010,Bono2017}. This learning occur concurrently with modifications of lateral interneuron weights which track changes in the output layer. Through the course of learning the network comes to a point where the novel top-down input is successfully predicted (Fig.~\ref{fig:learning_newinput}B,C).

\vspace*{-1mm}
\subsection{Network learns to solve a nonlinear regression task}
\vspace*{-1mm}

\begin{figure}[htb!]
  \centering
  \includegraphics[width=1\linewidth]{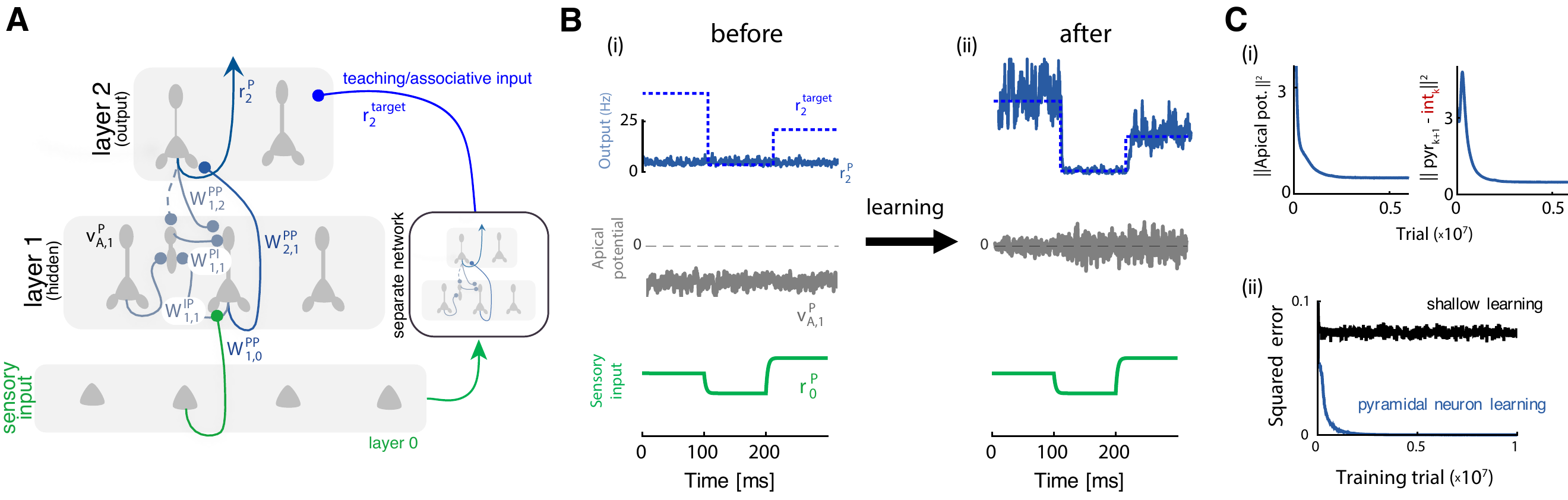}
  \caption{\textbf{Dendritic error microcircuit learns to solve a nonlinear regression task online and without phases.}
  (\textbf{A-C}) Starting from a random initial weight configuration, a 30-50-10 fully-connected network learns to approximate a nonlinear function (`separate network') from input-output pattern pairs.
  (\textbf{B}) Example firing rates for a randomly chosen output neuron ($r_2^\pn$, blue noisy trace) and its desired target imposed by the associative input ($r_2^\mathrm{target}$, blue dashed line), together with the voltage in the apical compartment of a hidden neuron ($v_{\apic,1}^\pn$, grey noisy trace) and the input rate from the sensory neuron ($r_0^\pn$, green). Traces are shown before (i) and after learning (ii). (\textbf{C}) Error curves for the full model and a shallow model for comparison.}
  \label{fig:learn-function-approximation}
\end{figure}

We now test the learning capabilities of the model on a nonlinear regression task, where the goal is to associate sensory input with the output of a separate multilayer network that transforms the same sensory input (Fig.~\ref{fig:learn-function-approximation}A). More precisely, a pyramidal neuron network of dimensions 30-50-10 (and 10 hidden layer interneurons) learns to approximate a random nonlinear function implemented by a held-aside feedforward network of dimensions 30-20-10. One teaching example consists of a randomly drawn input pattern $\mathbf{r}_0^\pn$ assigned to corresponding target $\mathbf{r}_2^\mathrm{trgt} = \phi(k_{2,1} \mathbf{W}^\mathrm{trgt}_{2,1} \, \phi(k_{1,0} \, \mathbf{W}^\mathrm{trgt}_{1,0} \, \mathbf{r}_0^\pn))$, with scale factors $k_{2,1} = 10$ and $k_{1,0} = 2$. Teacher network weights and input pattern entries are sampled from a uniform distribution $U(-1,1)$. We used a soft rectifying nonlinearity as the neuronal transfer function, $\phi(u) = \gamma \, \log(1 + \exp(\beta (u - \theta))$, with $\gamma = 0.1$, $\beta = 1$ and $\theta = 3$. This parameter setting led to neuronal activity in the nonlinear, sparse firing regime.

The network is initialized to a random initial synaptic weight configuration, with both pyramidal-pyramidal $\mathbf{W}^{\pn\pn}_{1,0}$, $\mathbf{W}^{\pn\pn}_{2,1}$, $\mathbf{W}^{\pn\pn}_{1,2}$ and pyramidal-interneuron weights $\mathbf{W}^{\intn\pn}_{1,1}$,  $\mathbf{W}^{\pn\intn}_{1,1}$ independently drawn from a uniform distribution. Top-down weight matrix $\mathbf{W}^{\pn\pn}_{1,2}$ is kept fixed throughout, in the spirit of feedback alignment \citep{Lillicrap2016}. Output layer teaching currents $\mathbf{i}_2^\pn$ are set so as to nudge $\mathbf{u}_2^\pn$ towards the teacher-generated $\mathbf{u}_2^\mathrm{trgt}$. Learning rates were manually chosen to yield best performance. Some learning rate tuning was required to ensure the microcircuit could track the changes in the bottom-up pyramidal-pyramidal weights, but we did not observe high sensitivity once the correct parameter regime was identified. Error curves are exponential moving averages of the sum of squared errors loss $\| \mathbf{r}_{2}^\pn - \mathbf{r}_{2}^\mathrm{trgt} \|^2$ computed after every example on unseen input patterns. Test error performance is measured in a noise-free setting ($\sigma = 0$). Plasticity induction terms given by Eqs.~\ref{eq:dW-PP}-\ref{eq:dW-PI} are low-pass filtered with time constant $\tau_w$ before being definitely consolidated, to dampen fluctuations; synaptic plasticity is kept on throughout. Plasticity and neuron model parameters are as defined above.

We let learning occur in continuous time without pauses or alternations in plasticity as input patterns are sequentially presented. This is in contrast to previous learning models that rely on computing activity differences over distinct phases, requiring temporally nonlocal computation, or globally coordinated plasticity rule switches  \citep{Hinton1988,OReilly1996,Xie2003,Scellier2017,Guerguiev2017}. Furthermore, we relaxed the bottom-up vs.~top-down weight symmetry imposed by backprop and kept the top-down weights $\mathbf{W}_{1,2}^{\pn\pn}$ fixed. Forward $\mathbf{W}_{1,2}^{\pn\pn}$ weights quickly aligned to $\sim\!45Âº$ of the feedback weights $\left(\mathbf{W}_{2,1}^{\pn\pn}\right)^T$ (see Fig.~\ref{fig:learn-self-predicting-state}), in line with the recently discovered feedback alignment phenomenon \citep{Lillicrap2016}. This simplifies the architecture, because top-down and interneuron-to-pyramidal synapses need not be changed. We set the scale of the top-down weights, apical and somatic conductances such that feedback and teaching inputs were strong, to test the model outside the weak feedback regime ($\lambda \to 0$) for which our SM theory was developed. Finally, to test robustness, we injected a weak noise current to every neuron.

Our network was able to learn this harder task  (Fig.~\ref{fig:learn-function-approximation}B), performing considerably better than a shallow learner where only hidden-to-output weights were adjusted (Fig.~\ref{fig:learn-function-approximation}C). Useful changes were thus made to hidden layer bottom-up weights. The self-predicting network state emerged throughout learning from a random initial configuration (see SM; Fig.~\ref{fig:learn-self-predicting-state}).

\vspace*{-1mm}
\subsection{Microcircuit network learns to classify handwritten digits}
\vspace*{-1mm}
\begin{figure}[htb!]
\floatbox[{\capbeside\thisfloatsetup{capbesideposition={right,top},capbesidewidth=5.5cm}}]{figure}[\FBwidth]
{\caption{\textbf{Dendritic error networks learn to classify handwritten digits.\label{fig:learn-mnist-classification}}
  (\textbf{A}) A network with two hidden layers learns to classify handwritten digits from the MNIST data set.
  (\textbf{B}) Classification error achieved on the MNIST testing set (blue; cf. shallow learner (black) and standard backprop\protect\footnotemark  (red)).}\label{fig:test}}
{\includegraphics[width=8cm]{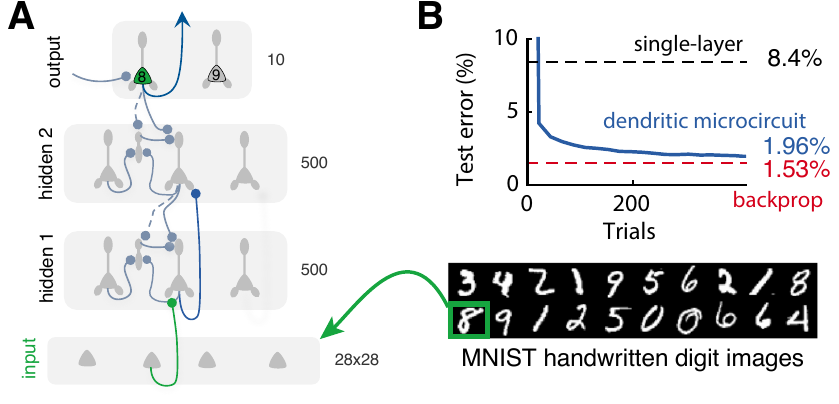}}

\end{figure}

Next, we turn to the problem of classifying MNIST handwritten digits. We wondered how our model would fare in this benchmark, in particular whether the prediction errors computed by the interneuron microcircuit would allow learning the weights of a hierarchical nonlinear network with multiple hidden layers. To that end, we trained a deeper, larger 4-layer network (with 784-500-500-10 pyramidal neurons,  Fig.~\ref{fig:learn-mnist-classification}A) by pairing digit images with teaching inputs that nudged the 10 output neurons towards the correct class pattern. We initialized the network to a random but self-predicting configuration where interneurons cancelled top-down inputs, rendering the apical compartments silent before training started. Top-down and interneuron-to-pyramidal weights were kept fixed.

Here for computational efficiency we used a simplified network dynamics where the compartmental potentials are updated only in two steps before applying synaptic changes. In particular, for each presented MNIST image, both pyramidal and interneurons are first initialized to their bottom-up prediction state \eqref{eq:Vbas}, $\mathbf{u}_k = \mathbf{v}_{\bas,k}$, starting from layer $1$ up to the top layer $N$. Output layer neurons are then nudged towards their desired target $\mathbf{u}_N^\mathrm{trgt}$, yielding updated somatic potentials $\mathbf{u}^\pn_N = (1 - \lambda_N) \, \mathbf{v}_{\bas,N} + \lambda_N \, \mathbf{u}^\mathrm{trgt}_N$. To obtain the remaining final compartmental potentials, the network is visited in reverse order, proceeding from layer $k=N-1$ down to $k=1$. For each $k$, interneurons are first updated to include top-down teaching signals, $\mathbf{u}_k^\intn = (1 - \lambda_I) \, \mathbf{v}_k^\intn + \lambda_I \, \mathbf{u}_{k+1}^\pn$; this yields apical compartment potentials according to \eqref{eq:Vapic}, after which we update hidden layer somatic potentials as a convex combination with mixing factor $\lambda_k$. The convex combination factors introduced above are directly related to neuron model parameters as conductance ratios. Synaptic weights are then updated according to Eqs.~\ref{eq:dW-PP}-\ref{eq:dW-PP-TD}. Such simplified dynamics approximates the full recurrent network relaxation in the deterministic setting $\sigma \to 0$, with the approximation improving as the top-down dendritic coupling is decreased, $g_\apic \to 0$.

We train the models on the standard MNIST handwritten image database, further splitting the training set into 55000 training and 5000 validation examples. The reported test error curves are computed on the 10000 held-aside test images. The four-layer network shown in Fig.~\ref{fig:learn-mnist-classification} is initialized in a self-predicting state with appropriately scaled initial weight matrices.  For our MNIST networks, we used relatively weak feedback weights, apical and somatic conductances (see SM) to justify our simplified approximate dynamics described above, although we found that performance did not appreciably degrade with larger values. To speed-up training we use a mini-batch strategy on every learning rule, whereby weight changes are averaged across 10 images before being applied. We take the neuronal transfer function $\phi$ to be a logistic function, $\phi(u) = 1/(1 + \exp(-u))$ and include a learnable threshold on each neuron, modelled as an additional input fixed at unity with a plastic weight. Desired target class vectors are 1-hot coded, with $r_N^\mathrm{trgt} \in \{0.1, 0.8\}$. During testing, the output is determined by picking the class label corresponding to the neuron with highest firing rate. We found the model to be relatively robust to learning rate tuning on the MNIST task, except for the rescaling by the inverse mixing factor to compensate for teaching signal dilution (see SM for the exact parameters).

The network was able to achieve a test error of 1.96\%, Fig.~\ref{fig:learn-mnist-classification}B, a figure not overly far from the reference mark of non-convolutional artificial neural networks optimized with backprop (1.53\%) and comparable to recently published results that lie within the range 1.6-2.4\% \citep{Lee2015,Lillicrap2016,Nokland2016}. The performance of our model also compares favorably to the 3.2\% test error reported by \cite{Guerguiev2017} for a two-hidden-layer network. This was possible despite the asymmetry of forward and top-down weights and at odds with exact backprop, thanks to a feedback alignment dynamics. Apical compartment voltages remained approximately silent when output nudging was turned off (data not shown), reflecting the maintenance of a self-predicting state throughout learning, which enabled the propagation of errors through the network. To further demonstrate that the microcircuit was able to propagate errors to deeper hidden layers, and that the task was not being solved by making useful changes only to the weights onto the topmost hidden layer, we re-ran the experiment while keeping fixed the pyramidal-pyramidal weights connecting the two hidden layers. The network still learned the dataset and achieved a test error of 2.11\%.

As top-down weights are likely plastic in cortex, we also trained a one-hidden-layer (784-1000-10) network where top-down weights were learned on a slow time-scale according to learning rule \eqref{eq:dW-PP-TD}. This inverse learning scheme is closely related to target propagation \citep{Bengio2014,Lee2015}. Such learning could play a role in perceptual denoising, pattern completion and disambiguation, and boost alignment beyond that achieved by pure feedback alignment \citep{Bengio2014}. Starting from random initial conditions and keeping all weights plastic (bottom-up, lateral and top-down) throughout, our network achieved a test classification performance of 2.48\% on MNIST. Once more, useful changes were made to hidden synapses, even though the microcircuit had to track changes in both the bottom-up and the top-down pathways.

\vspace{-0.0cm}
\section{Conclusions}
\vspace{-0.0cm}

Our work makes several predictions across different levels of investigation. Here we briefly highlight some of these predictions and related experimental observations. The most fundamental feature of the model is that distal dendrites encode error signals that instruct learning of lateral and bottom-up connections. While monitoring such dendritic signals during learning is challenging, recent experimental evidence suggests that prediction errors in mouse visual cortex arise from a failure to locally inhibit motor feedback~\citep{Zmarz:2016jo,Attinger:2017bo}, consistent with our model. Interestingly, the plasticity rule for apical dendritic inhibition, which is central to error encoding in the model, received support from another recent experimental study \citep{chiu2018input}.

A further implication of our model is that prediction errors occurring at a higher-order cortical area would imply also prediction errors co-occurring at earlier areas. Recent experimental observations in the macaque face-processing hierarchy support this \citep{Schwiedrzik:2017gh}.

Here we have focused on the role of a specific interneuron type (SST) as a feedback-specific interneuron. There are many more interneuron types that we do not consider in our framework. One such type are the PV (parvalbumin-positive) cells, which have been postulated to mediate a somatic excitation-inhibition balance \citep{Vogels2011,Froemke:2015kx} and competition \citep{Masquelier2007,Nessler:2013bu}.
These functions could in principle be combined with our framework in that PV interneurons may be involved in representing another type of prediction error (e.g., generative errors).

Humans have the ability to perform fast (e.g., one-shot) learning, whereas neural networks trained by backpropagation of error (or approximations thereof, like ours) require iterating over many training examples to learn. This is an important open problem that stands in the way of understanding the neuronal basis of intelligence. One possibility where our model naturally fits is to consider multiple subsystems (for example, the neocortex and the hippocampus) that transfer knowledge to each other and learn at different rates \citep{McClelland1995,Kumaran2016}.

Overall, our work provides a new view on how the brain may solve the credit assignment problem for time-continuous input streams by approximating the backpropagation algorithm, and bringing together many puzzling features of cortical microcircuits.

\subsubsection*{Acknowledgements}
The authors would like to thank Timothy P.~Lillicrap, Blake Richards, Benjamin Scellier and Mihai A.~Petrovici for helpful discussions. WS thanks Matthew Larkum for many inspiring discussions on dendritic processing. JS thanks Elena Kreutzer, Pascal Leimer and Martin T.~Wiechert for valuable feedback and critical reading of the manuscript.

This work has been supported by the Swiss National Science Foundation (grant 310030L-156863 of WS), the European Union’s Horizon 2020 Framework Programme for Research and Innovation under the Specific Grant Agreement No. 785907 (Human Brain Project), NSERC, CIFAR, and Canada Research Chairs.

\newpage
\section*{References}

{\def\section *#1{\small}

\bibliographystyle{apalike}
\bibliography{pyramidal}

}

\newpage
\setcounter{page}{1}

\section*{\large{Supplementary Material: Dendritic cortical microcircuits approximate the backpropagation algorithm}}

\setcounter{figure}{0} \renewcommand{\thefigure}{S\arabic{figure}}
\appendix

\subsection*{The dendritic cortical circuit learns to predict self-generated top-down input}

\begin{figure}[htb!]
  \centering
  \includegraphics[width=1\linewidth]{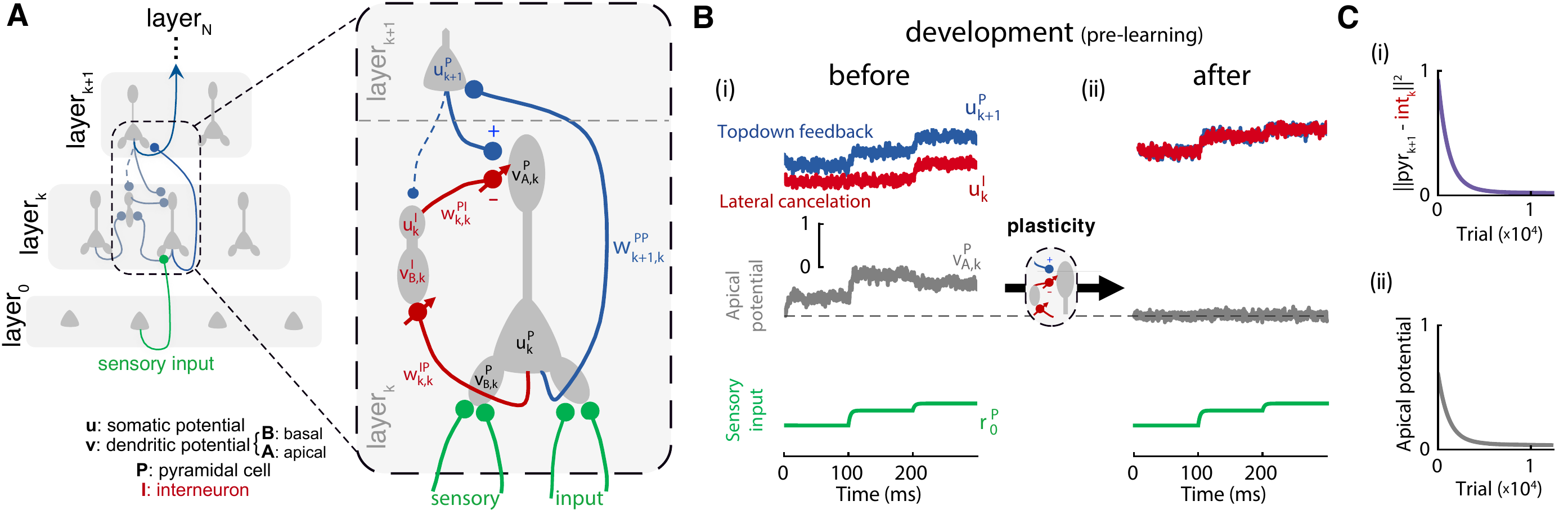}
  \caption{\textbf{Dendritic cortical circuit learns to predict self-generated top-down input.}
  (\textbf{A}) Illustration of multilayer network architecture. The network consists of an input layer $0$ (e.g., thalamic input), one or more intermediate (hidden) layers (represented by layer $k$ and layer $k\!+\!1$, which can be mapped onto primary and higher sensory layers) and an output layer $N$ (e.g., motor cortex) (left). Each hidden layer consists of a microcircuit with pyramidal cells and lateral inhibitory interneurons (e.g., SST cells) (right). Pyramidal cells consist of three compartments: a basal compartment (with voltage $\mathbf{v}_{\bas,k}^\pn$) that receives bottom-up input; an apical compartment (with voltage $\mathbf{v}_{\apic,k}^\pn$), where top-down input converges to; and a somatic compartment (with voltage $\mathbf{u}_k^\pn$), that integrates the basal and apical voltage. Interneurons receive input from lateral pyramidal cells onto their own basal dendrites (with voltage $\mathbf{v}_{\bas,k}^\intn$), integrate this input on their soma (with voltage $\mathbf{u}_k^\intn$) and project back to the apical compartments (with voltage $\mathbf{v}_{\apic,k}^\pn$) of same-layer pyramidal cells.
  (\textbf{B}) In a pre-learning developmental stage, the network learns to predict and cancel top-down feedback given randomly generated inputs. Only pyramidal-to-interneuron synapses ($\mathbf{W}^{\intn\pn}_{k,k}$) and interneuron-to-pyramidal synapses ($\mathbf{W}^{\pn\intn}_{k,k}$) are changed at that stage according to predictive synaptic plasticity rules (defined in Eqs.~\ref{eq:dW-IP} and \ref{eq:dW-PI}). Example voltage traces for a randomly chosen downstream neuron ($\mathbf{u}^\pn_{k+1}$) and a corresponding interneuron ($\mathbf{u}^\intn_k$), a pyramidal cell apical compartment ($\mathbf{v}_{\apic,k}^\pn$) and an input neuron ($\mathbf{u}_0^\pn$), before (i) and after (ii) development, for three consecutively presented input patterns. Once learning of the lateral synapses from and onto interneurons has converged, self-generated top-down signals are predicted by the network --- it is in a \emph{self-predicting state}. Here we use a concrete network with one hidden layer and 30-20-10 pyramidal neurons (input-hidden-output). Note that no desired targets are presented to the output layer (cf.~Fig.~\ref{fig:learning_newinput}); the network is solely driven by random inputs.
  (\textbf{C}) Lateral inhibition cancels top-down input. (i) Interneurons learn to match next-layer pyramidal neuron activity as their input weights $\mathbf{W}^{\intn\pn}_{k,k}$ adapt (see main text for details). (ii) Concurrently, learning of interneuron-to-pyramidal synapses ($\mathbf{W}^{\pn\intn}_{k,k}$) silences the apical compartment of pyramidal neurons, but pyramidal neurons remain active (cf.~B). This is a general effect, as the lateral microcircuit learns to predict and cancel the expected top-down input for every random pattern.}
  \label{fig:learn-self-predicting-state}
\end{figure}

The microcircuit model introduced in the main text is key to encode and backpropagate errors across the network. Here, we illustrate how synaptic plasticity of lateral interneuron connections establishes a network regime, which we term \emph{self-predicting}, whereby lateral input cancels the self-generated top-down feedback, effectively silencing apical dendrites. For this reason, SST cells are functionally inhibitory and are henceforth referred to as interneurons. Crucially, when the circuit is in this so-called self-predicting state, presenting a novel external signal at the output layer gives rise to top-down activity that cannot be explained away by the interneuron circuit. Below we show that these apical mismatches between top-down and lateral input constitute backpropagated, neuron-specific errors that drive plasticity on the forward weights to the hidden pyramidal neurons.

Learning to predict the feedback signals involves adapting both weights from and to the lateral interneuron circuit. Consider a network that is driven by a succession of sensory input patterns (Fig.~\ref{fig:learn-self-predicting-state}B, bottom row). Learning to cancel the feedback input is divided between both the weights from pyramidal cells to interneurons, $\mathbf{W}_{k,k}^{\intn\pn}$, and from interneurons to pyramidal cells, $\mathbf{W}_{k,k}^{\pn\intn}$.

First, due to the somatic teaching feedback, learning of the $\mathbf{W}_{k,k}^{\intn\pn}$ weights leads interneurons to better reproduce the activity of the respective higher layer $k\!+\!1$ (Fig.~\ref{fig:learn-self-predicting-state}B (i)). A failure to reproduce layer $k\!+\!1$ activity generates an internal prediction error at the dendrites of the interneurons, which triggers synaptic plasticity (as defined by Eq.~\ref{eq:dW-IP}) that corrects for the wrong dendritic prediction and eventually leads to a faithful tracing of the upper layer activity by the lower layer interneurons  (Fig.~\ref{fig:learn-self-predicting-state}B (ii)). The mathematical analysis (see section below, Eq.~\ref{eq:expected-dW-IP}) shows that the plasticity rule \eqref{eq:dW-IP} makes the inhibitory population implement the same function of the layer-$k$ pyramidal cell activity as done by the layer--($k\!+\!1$) pyramidal neurons. Thus, the interneurons will learn to mimic the layer--($k\!+\!1$) pyramidal neurons (Fig.~\ref{fig:learn-self-predicting-state}Ci).

Second, as the interneurons mirror upper layer activity, inter-to-pyramidal neuron synapses within the same layer ($\mathbf{W}_{k,k}^{\pn\intn}$, Eq.~\ref{eq:dW-PI}) successfully learn to cancel the top-down input to the apical dendrite (Fig.~\ref{fig:learn-self-predicting-state}Cii), independently of the actual input stimulus that drives the network. By doing so, the inter-to-pyramidal neuron weights $\mathbf{W}_{k,k}^{\pn\intn}$ learn to mirror the top-down weights onto the lower layer pyramidal neurons. The learning of the weights onto and from the interneurons works in parallel: as the interneurons begin to predict the activity of pyramidal cells in layer $k\!+\!1$, it becomes possible for the plasticity at interneuron-to-pyramidal synapses (Eq.~\ref{eq:dW-PI}) to find a synaptic weight configuration which precisely cancels the top-down feedback (see also Eq.~\ref{eq:expected-dW-PI} below). At this stage, every pattern of activity generated by the hidden layers of the network is explained by the lateral circuitry, Fig.~\ref{fig:learn-self-predicting-state}C~(ii). Importantly, once learning of the lateral interneurons has converged, the apical input cancellation occurs irrespective of the actual bottom-up sensory input. Therefore, interneuron synaptic plasticity leads the network to a \emph{self-predicting state}.

\begin{figure}[htb!]
  \centering
  \includegraphics[width=1\textwidth]{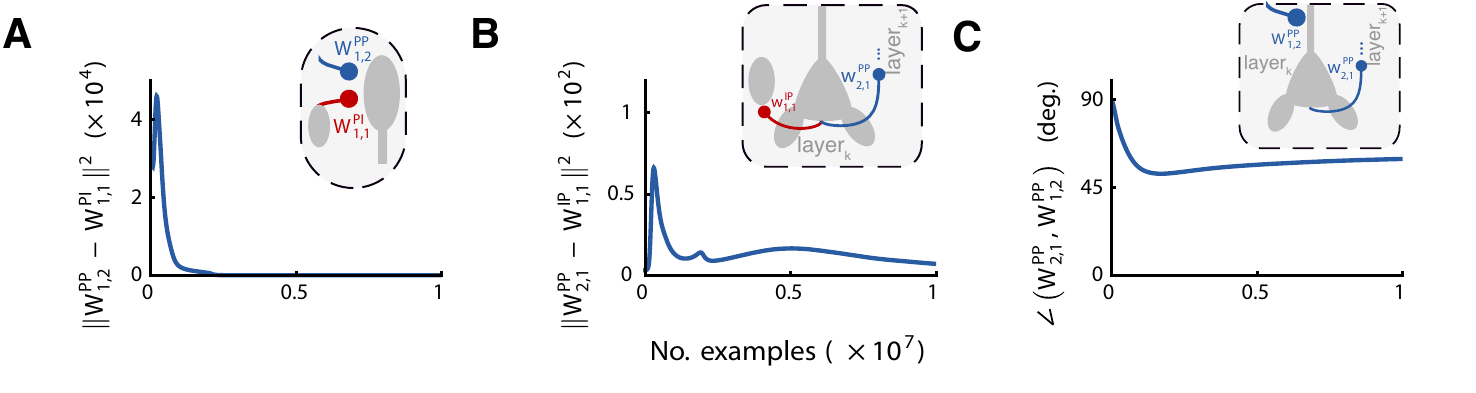}
  \caption{\textbf{Emergence and maintenance of a self-predicting network state while learning a target function.} (\textbf{A}, \textbf{B}) Starting from random initial conditions (see Fig.~\ref{fig:learn-function-approximation}), co-evolving bottom-up pyramidal-pyramidal and lateral microcircuit pyramidal-interneuron synaptic weights lead the network to a self-predicting state. To quantify the approximation error, we used the squared Frobenius matrix norm. Pyramidal-to-interneuron $\mathbf{W}_{1,1}^{\intn\pn}$ and apical-targetting weights $\mathbf{W}_{1,1}^{\pn\intn}$ approach ideal values (cf. supplementary mathematical analysis, and Fig.~\ref{fig:learn-self-predicting-state} and Fig.~\ref{fig:learning_newinput}), allowing the backpropagation of output errors to layer 1 neurons. This state is maintained throughout, as bottom-up weights learn the target function (Fig.~\ref{fig:learn-function-approximation}C). (\textbf{C}) Quickly after learning starts, bottom-up and top-down pyramidal-pyramidal weights align, a phenomenon known as feedback alignment \citep{Lillicrap2016}; by virtue of simultaneous pyramidal and interneuron synaptic plasticity the network effectively learns how to backpropagate errors. \label{fig:self-predicting-supp}}
\end{figure}

We propose that the emergence of this state could occur during development, consistent with experimental findings \citep{Dorrn:2010hu,Froemke:2015kx}. Starting from a cross-layer self-predicting configuration helps speed-up learning of specific tasks, but is not essential. Indeed, we were able to train a nonlinear regression model (cf.~Fig.~\ref{fig:learn-function-approximation}) and an MNIST network starting from random conditions. Appropriate tuning of learning rates quickly led the network to a self-predicting state, which unlocked learning of the task, see Fig.~\ref{fig:self-predicting-supp}.

\newpage

\subsection*{Supplementary data}
Below we detail the model parameters used to generate the figures presented in the paper.

\vspace{1em}
\textbf{Fig.~\ref{fig:learn-self-predicting-state} details}. The parameters for the compartmental model neuron were: $g_\apic = 0.8$, $g_\bas = g_\dnd = 1.0$, $g_\leak = 0.1$. Interneuron somatic teaching conductances were balanced to yield overall nudging strength $g_\som = 0.8$. Initial weight matrix entries were independently drawn from a uniform distribution $U(-1, 1)$. We used a soft rectifying transfer function $\phi(u) = \log(1 + \exp(u))$. We chose background activity levels of $\sigma=0.1$. The learning rates were set as $\eta^{\intn\pn}_{1,1} = 0.0002375$ and $\eta^{\pn\intn}_{1,1} = 0.0005$.

Input patterns were smoothly transitioned by low-pass filtering $\mathbf{u}_0^\pn$ with time constant $\tau_0 = 3$. A transition between patterns was triggered every 100 ms. Weight changes were low pass filtered with time constant $\tau_w = 30$. The dynamical equations were solved using Euler's method with a time step of 0.1, which resulted in 1000 integration time steps per pattern.

\vspace{1em}
\textbf{Fig.~\ref{fig:learning_newinput} details}. We used learning rates $\eta_{1,0}^{\pn\pn} = \eta_{1,1}^{\intn\pn} = 0.0011875$ and $\eta_{2,1}^{\pn\pn} = 0.0005$. Remaining parameters as used for Fig.~\ref{fig:learn-self-predicting-state}.

\vspace{1em}
\textbf{Fig.~\ref{fig:learn-function-approximation} details}. Initial forward  and pyramidal-interneuron weights were drawn independently from a uniform distribution $U(-0.1, 0.1)$. The network learned under a background noise level of $\sigma = 0.3$. The learning rates were $\eta^{\intn\pn}_{1,1} = 0.0011875$, $\eta^{\pn\intn}_{1,1} = 0.0059375$, $\eta^{\pn\pn}_{1,0} = 0.0011875$, $\eta^{\pn\pn}_{2,1} = 0.0005$. Weight matrix $\mathbf{W}^{\pn\pn}_{1,2}$ was kept fixed, so the model relied on a feedback alignment mechanism to learn. Remaining  parameters as used for Fig.~\ref{fig:learn-self-predicting-state}.

\vspace{1em}
\textbf{Fig.~\ref{fig:learn-mnist-classification} details}. We chose mixing factors $\lambda_3 = \lambda_I = 0.1$ and $\lambda_1 = \lambda_2 = 0.3$. Forward learning rates were $\eta_{3,2}^{\pn\pn} = 0.001/\lambda_3$, $\eta_{2,1}^{\pn\pn} = \eta_{3,2}^{\pn\pn}/\lambda_2$, $\eta_{1,0}^{\pn\pn} = \eta_{2,1}^{\pn\pn}/\lambda_1$. Lateral learning rates were $\eta_{2,2}^{\intn\pn} = 2 \eta_{3,2}^{\pn\pn}$ and $\eta_{1,1}^{\intn\pn} = 2 \eta_{2,1}^{\pn\pn}$. Initial forward weights were drawn at random from a uniform distribution $U(-0.1, 0.1)$, and the remaining weights from $U(-1,1)$.

\newpage
\subsection*{Supplementary analysis}
In this supplementary note we present a set of mathematical results concerning the network and plasticity model described in the main text.

To proceed analytically we make a number of simplifying assumptions. Unless noted otherwise, we study the network in a deterministic setting and consider the limiting case where lateral microcircuit synaptic weights match the corresponding forward weights:
\begin{align}
  \label{eq:ideal-wpi} \mathbf{W}^{\pn \intn}_{k, k} &= - \mathbf{W}_{k, k+1}^{\pn \pn} \equiv \mathbf{W}^{\pn \intn*}_{k,k}\\
  \label{eq:ideal-wip}\mathbf{W}^{\intn \pn}_{k, k} &= \frac{g_\bas + g_\leak}{g_\bas + g_\apic + g_\leak} \mathbf{W}_{k+1, k}^{\pn \pn} \equiv \mathbf{W}^{\intn \pn*}_{k,k},
\end{align}
The particular choice of proportionality factors, which depend on the neuron model parameters, is motivated below. Under the above configuration, the network becomes self-predicting.

To formally relate the encoding and propagation of errors implemented by the inhibitory microcircuit to the backpropagation of errors algorithm from machine learning, we consider the limit where top-down input is weak compared to the bottom-up drive. This limiting case results in error signals that decrease exponentially with layer depth, but allows us to proceed analytically.

We further assume that the top-down weights converging to the apical compartments are equal to the corresponding forward weights, $\mathbf{W}^{\pn\pn}_{k, k+1} = \left(\mathbf{W}^{\pn\pn}_{k+1, k} \right)^T$. Such weight symmetry is not essential for successful learning in a broad range of problems, as demonstrated in the main simulations and as observed before \citep{Lee2015,Lillicrap2016,Nokland2016}. It is, however, required to frame learning as a gradient descent procedure. Furthermore, in the analyses of the learning rules, we assume that synaptic changes take place at a fixed point of the neuronal dynamics; we therefore consider discrete-time versions of the plasticity rules. This approximates the continuous-time plasticity model as long as changes in the inputs are slow compared to the neuronal dynamics.

For convenience, we will occasionally drop neuron type indices and refer to bottom-up weights $\mathbf{W}_{k+1, k}$ and to top-down weights $\mathbf{W}_{k, k+1}$. Additionally, we assume without loss of generality that the dendritic coupling conductance for interneurons is equal to the basal dendritic coupling of pyramidal neurons, $g_\dnd = g_\bas$. Finally, whenever it is useful to distinguish whether output layer nudging is turned off, we use superscript `$-$'.

\vspace{1em}
\textbf{Interneuron activity in the self-predicting state.}
Following \citet{Urbanczik2014}, we note that steady state interneuron somatic potentials can be expressed as a convex combination of basal dendritic and pyramidal neuron potentials that are provided via somatic teaching input:
\begin{equation}
\label{eq:intn-convcomb}
  \mathbf{u}_k^{\intn} =  \frac{g_\bas}{g_\leak + g_\bas + g_\som} \, \mathbf{v}_{k}^{\intn} + \frac{g_\som}{g_\leak + g_\bas + g_\som} \, \mathbf{u}_{k+1}^{\pn} = (1 - \lambda) \,  \hat{\mathbf{v}}_{k}^{\intn} + \lambda \, \mathbf{u}_{k+1}^{\pn},
\end{equation}
with $g_\bas$ and $g_\leak$ the effective dendritic transfer and leak conductances, respectively, and $g_\som$ the total excitatory and inhibitory teaching conductance. In the equation above, $\hat{\mathbf{v}}_{k}^{\intn} = \frac{g_\bas}{g_\leak + g_\bas} \mathbf{v}_{k}^{\intn}$ is the interneuron dendritic prediction (cf.~Eq.~\ref{eq:dW-IP}), and $\lambda \equiv \frac{g_\som}{g_\leak + g_\bas + g_\som} \in [0,1[$ is a mixing factor which controls the nudging strength for the interneurons. In other words, the current prediction $\hat{\mathbf{v}}_{k}^{\intn}$ and the teaching signal are averaged with coefficients determined by normalized conductances. We will later consider the weak nudging limit of $\lambda \to 0$.

The relation $\hat{\mathbf{v}}_{k}^{\intn} = \hat{\mathbf{v}}_{\bas,k+1}^{\pn}$ holds when pyramidal-to-interneuron synaptic weights are equal to pyramidal-pyramidal forward weights, up to a scale factor: $\mathbf{W}^{\intn\pn}_{k,k} = \frac{g_\leak + g_\bas}{g_\leak + g_\bas + g_\apic} \mathbf{W}^{\pn\pn}_{k+1,k}$, which simplifies to $\mathbf{W}^{\intn\pn}_{N-1,N-1} = \mathbf{W}^{\pn\pn}_{N,N-1}$ for the last layer where $g_\apic=0$ (to reduce clutter, we use the slightly abusive notation whereby $g_\apic$ should be understood to be zero when referring to output layer neurons). This is the reason for the particular choice of ideal pyramidal-to-interneuron weights presented in the preamble. The network is then internally consistent, in the sense that the interneurons predict the model's own predictions, held by pyramidal neurons.

\vspace{1em}
\textbf{Bottom-up predictions in the absence of external nudging.}
We first study the situation where the input pattern $\mathbf{r}_0$ is stationary and the output layer teaching input is disabled, $\mathbf{i}^\pn_{N} = 0$. We show that the fixed point of the network dynamics is a state where somatic voltages are equal to basal voltages, up to a dendritic attenuation factor. In other words, the network effectively behaves as if it were feedforward, in the sense that it computes the same function as the corresponding network with equal bottom-up but no top-down or lateral connections.

Specifically, in the absence of external nudging (indicated by the $-$ in the superscript), the somatic voltages of pyramidal and interneuron are given by the bottom-up dendritic predictions,
\begin{align}
  \label{eq:FF-state}
  \mathbf{u}^{\pn,-}_k = \hat{\mathbf{v}}_{\bas,k}^{\pn,-} &\equiv \frac{g_\bas}{g_\leak + g_\bas + g_\apic} \, \mathbf{W}^{\pn \pn}_{k, k-1} \, \phi(\hat{\mathbf{v}}_{\bas, k-1}^{\pn,-})\\
  \mathbf{u}^{\intn,-}_k = \hat{\mathbf{v}}_{k}^{\intn,-} &\equiv \frac{g_\bas}{g_\leak + g_\bas} \, \mathbf{W}^{\intn \pn}_{k, k} \, \phi(\hat{\mathbf{v}}_{\bas, k}^{\pn,-}).
\end{align}

To show that Eq.~\ref{eq:FF-state} describes the state of the network, we start at the output layer and set Eq.~\ref{eq:dUPdt} to zero. Because nudging is turned off, we observe that $\mathbf{u}_N^{\pn}$ is equal to $\hat{\mathbf{v}}_{\bas, N}^{\pn, -}$ if layer $N-1$ also satisfies $\mathbf{u}_{N-1}^{\pn} = \hat{\mathbf{v}}_{\bas, N-1}^{\pn, -}$. The same recursively applies to the hidden layer below when its apical voltage vanishes, $\mathbf{v}_{\apic, N-1}^\pn = 0$. Now we note that at the fixed point the interneuron cancels the corresponding pyramidal neuron, due to the assumption that the network is in a self-predicting state, which yields $\mathbf{u}_{N-1}^{\intn} = \mathbf{u}_{N}^{\pn}$. Together with the fact that $\mathbf{W}^{\pn \intn}_{N-1, N-1} = - \mathbf{W}_{N-1, N}^{\pn \pn}$, we conclude that the interneuron contribution to the apical compartment cancels the top-down pyramidal neuron input, yielding the required condition $\mathbf{v}_{\apic, N-1}^\pn = 0$.

The above argument can be iterated down to the input layer, where activity is constant, and we arrive at Eq.~\ref{eq:FF-state}.

\vspace{1em}
\textbf{Zero plasticity induction in the absence of nudging.} In view of Eq.~\ref{eq:FF-state}, which states that in the absence of external nudging the somatic voltages correspond to the basal predictions, no synaptic changes are induced in basal synapses on the pyramidal and interneurons as defined by the plasticity rules \eqref{eq:dW-PP} and \eqref{eq:dW-IP}, respectively. Similarly, the apical voltages are equal to rest, $\mathbf{v}^{\pn,-}_{\apic,k}=\mathbf{v}_\rest$, when the top-down input is fully predicted, and no synaptic plasticity is induced in the inter-to-pyramidal neuron synapses, see \eqref{eq:dW-PI}. When noisy background currents are present, the average prediction error is zero, while momentary fluctuations will still trigger plasticity. Note that the above holds when the dynamics is away from equilibrium, under the additional constraint that the integration time constant of interneurons matches that of pyramidal neurons.

\vspace{1em}
\textbf{Recursive prediction error propagation.}
Prediction errors arise in the model whenever lateral interneurons cannot fully explain top-down input, leading to a deviation from baseline in apical dendrite activity. Here, we look at the network steady state equations for a stationary input pattern $\mathbf{r}_0$ and derive an iterative relationship which establishes the propagation across the network of prediction mismatches originating downstream. The following compartmental potentials are thus evaluated at a fixed point of the neuronal dynamics.

Under the assumption \eqref{eq:ideal-wpi} of matching interneuron-to-pyramidal top-down weights, apical compartment potentials simplify to
\begin{equation}
\label{eq:v-apic}
  \mathbf{v}^\pn_{\apic,k} = \mathbf{W}_{k, k+1} \left[\phi(\mathbf{u}^\pn_{k+1}) - \phi(\mathbf{u}^\intn_{k})\right] = \mathbf{W}_{k, k+1} \, \mathbf{e}_{k+1},
\end{equation}
where we introduced error vector $\mathbf{e}_{k+1}$ defined as the difference between pyramidal and interneuron firing rates. Such deviation can be intuitively understood as an layer-wise interneuron prediction mismatch, being zero when interneurons perfectly explain pyramidal neuron activity. We now evaluate this difference vector at a fixed point to obtain a recurrence relation that links consecutive layers.

The steady-state somatic potentials of hidden pyramidal neurons are given by
\begin{align}
\label{eq:pn-star}
  \mathbf{u}_{k}^{\pn} &=  \frac{g_\bas}{g_\leak + g_\bas + g_\apic} \, \mathbf{v}_{\bas,k}^{\pn} + \frac{g_\apic}{g_\leak + g_\bas + g_\apic} \, \mathbf{v}_{\apic,k}^{\pn} = \hat{\mathbf{v}}_{\bas,k}^{\pn} + \lambda \, \mathbf{v}_{\apic,k}^{\pn}\nonumber\\
  &= \hat{\mathbf{v}}_{\bas,k}^{\pn} + \lambda \, \mathbf{W}_{k, k+1} \, \mathbf{e}_{k+1}.
\end{align}
To shorten the following, we assumed that the apical attenuation factor is equal to the interneuron nudging strength $\lambda$. As previously mentioned, we proceed under the assumption of weak feedback, $\lambda$ small. As for the corresponding interneurons, we insert Eq.~\ref{eq:pn-star} into Eq.~\ref{eq:intn-convcomb} and note that when the network is in a self-predicting state we have $\hat{\mathbf{v}}_{k-1}^{\intn} = \hat{\mathbf{v}}_{\bas,{k}}^{\pn}$, yielding
\begin{equation}
\label{eq:intn-star}
  \mathbf{u}_{k-1}^{\intn} =  (1 - \lambda) \,  \hat{\mathbf{v}}_{\bas,k}^{\pn} + \lambda \left(\hat{\mathbf{v}}_{\bas,k}^{\pn} + \lambda \, \mathbf{v}_{\apic,k}^{\pn} \right) = \hat{\mathbf{v}}_{\bas,k}^{\pn} +  \lambda^2 \, \mathbf{v}_{\apic,k}^{\pn}.
\end{equation}

Using the identities \eqref{eq:pn-star} and \eqref{eq:intn-star}, we now expand to first order the difference vector $\mathbf{e}_{k}$ around $\hat{\mathbf{v}}_{\bas,k}^{\pn}$ as follows
\begin{align}
  \mathbf{e}_{k} &= \phi(\mathbf{u}_{k}^\pn) - \phi(\mathbf{u}_{k-1}^\intn) = \lambda  \, \mathbf{D}_{k} \, \mathbf{v}_{\apic,k}^{\pn} + \mathcal{O}\!\left(\lambda^2 \, \| \mathbf{v}_{\apic,k}^\pn \|\right).
\end{align}
Matrix $\mathbf{D}_{k}$ is a diagonal matrix with diagonal equal to $\phi^\prime(\hat{\mathbf{v}}_{\bas,k}^{\pn})$, i.e., whose $i$-th element reads $\frac{d\phi}{dv} (\hat{v}_{\bas,k,i}^{\pn})$. It contains the derivative of the neuronal transfer function $\phi$ evaluated component-wise at the bottom-up predictions $\hat{\mathbf{v}}_{\bas,k+1}^{\pn}$. Recalling Eq.~\ref{eq:v-apic}, we obtain a recurrence relation
\begin{align}
  \mathbf{e}_{k} &= \lambda \, \mathbf{D}_{k} \, \mathbf{W}_{k, k+1} \, \mathbf{e}_{k+1} + \mathcal{O}\!\left(\lambda^2 \, \| \mathbf{W}_{k,k+1} \, \mathbf{e}_{k+1} \|\right).
\end{align}

Finally, last layer pyramidal neurons provide the initial condition by being directly nudged towards the desired target  $\mathbf{u}_N^\text{trgt}$. Their membrane potentials can be written as
\begin{equation}
  \mathbf{u}_{N}^{\pn} = (1 - \lambda) \, \hat{\mathbf{v}}_{\bas,N}^{\pn} + \lambda \, \mathbf{u}_N^\text{trgt},
\end{equation}
and this gives an estimate for the error in the output layer of the form
\begin{equation}
  \label{eq:delta-K}
  \mathbf{e}_{N} = \lambda \, \mathbf{D}_{N} \left(\mathbf{u}_N^\text{trgt} - \hat{\mathbf{v}}_{\bas,N}^{\pn}\right) + \mathcal{O}\!\left(\lambda^2 \, \|\mathbf{u}_N^\text{trgt} - \hat{\mathbf{v}}_{\bas,N}^{\pn} \|\right) ,
\end{equation}
where for simplicity we took the same mixing factor $\lambda$ for pyramidal output and interneurons. Then, for an arbitrary layer, assuming that the synaptic weights and the remaining fixed parameters do not scale with $\lambda$, we arrive at
\begin{equation}
  \label{eq:delta-star}
  \mathbf{e}_{k} = \lambda^{N-k+1} \, \left(\prod_{l=k}^{N-1} \mathbf{D}_{l} \, \mathbf{W}_{l, l+1} \right) \, \mathbf{D}_{N} \left(\mathbf{u}_N^\text{trgt} - \hat{\mathbf{v}}_{\bas,N}^{\pn}\right) + \mathcal{O}(\lambda^{N-k+2}).
\end{equation}

Thus, steady state potentials of apical dendrites (cf.~Eq.~\ref{eq:v-apic}) recursively encode neuron-specific prediction errors that can be traced back to a mismatch at the output layer.

\vspace{1em}
\textbf{Learning as approximate error backpropagation.}
In the previous section we found that neurons implicitly carry and transmit error information across the network. We now show how the proposed synaptic plasticity model, when applied at a steady state of the neuronal dynamics, can be recast as an approximate gradient descent learning procedure.

More specifically, we compare our model against learning through backprop \citep{Rumelhart1986} or approximations thereof \citep{Lee2015,Lillicrap2016} the weights of the feedfoward multilayer network obtained by removing interneurons and top-down connections from the intact network. For this reference model, the activations $\mathbf{u}_k^-$ are by construction equal to the bottom-up predictions obtained in the full model when output nudging is turned off, $\mathbf{u}_k^- \equiv \hat{\mathbf{v}}_{\bas,k}^{\pn,-}$, cf.~Eq.~\ref{eq:FF-state}. Thus, optimizing the weights in the feedforward model is equivalent to optimizing the predictions of the full model.

We now assume that $\phi$ is monotonically increasing and define the loss function
\begin{equation}
  \mathcal{L} \! \left(\mathbf{u}_{N}^-, \mathbf{u}_{N}^\text{trgt} \right) = - \sum_{i=1}^{N_N} \int_0^{u_{N,i}^-} \phi\left((1-\lambda) \, \nu + \lambda \, u^\text{trgt}_{N,i}\right) - \phi(\nu) \, d\nu,
\end{equation}
where $N_N$ denotes the number of output neurons. $\mathcal{L}$ can be thought of as the multilayer, multi-output unit analogue of the loss function optimized by the single neuron model \citep{Urbanczik2014}, where it stems directly from the particular chosen form of the learning rule \eqref{eq:dW-PP}. The nudging strength parameter $\lambda \in [0, 1[$ allows controlling the mixing with the target and can be understood as an additional learning rate parameter. Albeit unusual in form, function $\mathcal{L}$ imposes a cost similar to an ordinary squared error loss. Importantly, it has a minimum when $\mathbf{u}^{-}_N = \mathbf{u}_N^\text{trgt}$ and it is lower bounded. Furthermore, it is differentiable with respect to compartmental voltages (and synaptic weights). It is therefore suitable for gradient descent optimization. As a side remark, $\mathcal{L}$ integrates to a quadratic function when $\phi$ is linear.

Gradient descent proceeds by changing synaptic weights according to
\begin{equation}
\Delta \mathbf{W}_{k,k-1} = - \eta \, \frac{\partial \mathcal{L}}{\partial \mathbf{W}_{k,k-1}}.
\end{equation}
The required partial derivatives can be efficiently computed by the backpropagation of errors algorithm. For the network architecture we study, this yields a learning rule of the form
\begin{equation}
  \label{eq:Delta-W-bp}
   \Delta \mathbf{W}_{k,k-1}^\text{bp} = \eta \, \mathbf{e}_{k}^- \, \phi(\mathbf{u}^-_{k-1})^T.
\end{equation}
The error factor $\mathbf{e}_{k}^-$ can be expressed recursively as follows:
\begin{equation}
  \mathbf{e}_{k}^- =
  \begin{cases}
    \phi \! \left((1-\lambda) \, \mathbf{u}_N^- + \lambda \, \mathbf{u}_N^\text{trgt} \right) - \phi \!\left(\mathbf{u}_N^-\right) & \text{ if } k = N,\\
    \mathbf{D}^-_k \, \mathbf{W}^T_{k+1,k} \, \mathbf{e}^-_{k+1} & \text{ otherwise,}
  \end{cases}
\end{equation}
ignoring constant factors that depend on conductance ratios, which can be dealt with by redefining learning rates or backward pass weights. As in the previous section, matrix $\mathbf{D}_k^-$ is a diagonal matrix, with diagonal equal to $\phi^\prime(\mathbf{u}^-_k)$.

We first compare the fixed point equations of the original network to the feedforward activations of the reference model. Starting from the bottom most hidden layer, using Eqs.~\ref{eq:v-apic}, \ref{eq:pn-star} and \ref{eq:delta-star}, we notice that $\mathbf{u}_1^{\pn} = \mathbf{u}_1^{-} + \lambda \, \mathbf{v}_{\apic,1}^{\pn} = \mathbf{u}_1^{-} + \mathcal{O}(\lambda^N)$, as the bottom-up input is the same in both cases. Inserting this into second hidden layer steady state potentials and linearizing the neuronal transfer function gives $\mathbf{u}_2^{\pn} = \mathbf{u}_2^{-} + \lambda \, \mathbf{v}_{\apic,2}^{\pn} + \mathcal{O}(\lambda^{N}) =  \mathbf{u}_2^{-} + \mathcal{O}(\lambda^{N-1})$. This can be repeated and for an arbitrary layer and neuron type we find
\begin{align}
  \label{eq:pn-minus-star} \mathbf{u}_k^{\pn} &= \mathbf{u}_k^-  + \lambda \, \mathbf{v}_{\apic,k}^{\pn} + \mathcal{O}(\lambda^{N-k+2}) = \mathbf{u}_k^- + \mathcal{O}(\lambda^{N-k+1})\\
  \label{eq:intn-minus-star}\mathbf{u}_{k-1}^{\intn} &= \mathbf{u}_k^- + \mathcal{O}(\lambda^{N-k+2}).
\end{align}
Writing Eq.~\ref{eq:pn-minus-star} in the first form emphasizes that the apical contributions dominate $\mathcal{O}(\lambda \, \mathbf{v}_{\apic,k}^{\pn}) = \mathcal{O}(\lambda^{N-k+1})$ the bottom-up corrections, which are of order $\mathcal{O}(\lambda^{N-k+2})$.

Next, we prove that up to a factor and to first order the apical term in Eq.~\ref{eq:pn-minus-star} represents the backpropagated error in the feedforward network, $\mathbf{e}_{k}^-$. Starting from the topmost hidden layer apical potentials, we reevaluate difference vector \eqref{eq:delta-K} using \eqref{eq:pn-minus-star}. Linearization of the neuronal transfer function gives
\begin{equation}
  \mathbf{v}_{\apic,N-1}^{\pn} =  \lambda \, \mathbf{W}_{N-1,N} \, \mathbf{D}^-_N \left(\mathbf{u}_N^\text{trgt} - \mathbf{u}_N^- \right) + \mathcal{O}(\lambda^2).
\end{equation}
Inserting the expression above into Eq.~\ref{eq:pn-minus-star} and using Eq.~\ref{eq:intn-minus-star} the apical compartment potentials at layer $N-1$ can then be recomputed. This procedure can be iterated until the input layer is reached. In general form, somatic membrane potentials at hidden layer $k$ can be expressed as
\begin{align}
  \mathbf{u}^{\pn}_k &= \mathbf{u}_k^- + \lambda \, \mathbf{v}_{\apic,k}^{\pn} + \mathcal{O}(\lambda^{N-k+2})\\
  &= \mathbf{u}_k^- + \lambda^{N-k+1} \, \mathbf{W}_{k,k+1} \, \left(\prod_{l=k+1}^{N-1} \mathbf{D}_{l}^- \, \mathbf{W}_{l, l+1} \right) \mathbf{D}_{N}^- \left(\mathbf{u}_N^\text{trgt} - \mathbf{u}_N^- \right) + \mathcal{O}(\lambda^{N-k+2}).
\end{align}
This equation shows that, to leading order of $\lambda$, hidden neurons mix and propagate forward purely bottom-up predictions with top-down errors that are computed at the output layer and spread backwards.

We are now in position to compare model synaptic weight updates to the ones prescribed by backprop. Output layer updates are exactly equal by construction, $\Delta \mathbf{W}_{N,N-1} = \Delta \mathbf{W}^\text{bp}_{N,N-1}$. For pyramidal-to-pyramidal neuron synapses from hidden layer $k-1$ to layer $k$, we obtain
\begin{align}
  \Delta \mathbf{W}_{k,k-1} &= \eta_{k,k-1} \left[ \phi(\mathbf{u}_{k}^{\pn}) - \phi(\hat{\mathbf{v}}_{\bas,k}^{\pn}) \right] \left(\mathbf{r}_{k-1}^{\pn}\right)^T \nonumber\\
  &= \eta_{k,k-1} \left[\phi\left(\mathbf{u}_{k}^- +  \lambda \, \mathbf{v}_{\apic,k}^{\pn} + \mathcal{O}(\lambda^{N-k+2})\right) - \phi(\mathbf{u}_{k}^-)\right] \left(\mathbf{r}_{k-1}^- + \mathcal{O}(\lambda^{N-k+2}) \right)^T \nonumber\\
  \label{eq:Delta-W-weak} &= \eta_{k,k-1} \lambda^{N-k+1} \left(\prod_{l=k}^{N-1} \mathbf{D}_{l}^- \, \mathbf{W}_{l, l+1} \right) \mathbf{D}_{N}^- \left(\mathbf{u}_N^\text{trgt} - \mathbf{u}_N^- \right) \left(\mathbf{r}_{k-1}^-\right)^T + \mathcal{O}(\lambda^{N-k+2}),
\end{align}
while backprop learning rule \eqref{eq:Delta-W-bp} can be written as
\begin{align}
  \Delta \mathbf{W}_{k,k-1}^\text{bp} &= \eta \, \lambda \, \left(\prod_{l=k}^{N-1} \mathbf{D}_{l}^- \, \mathbf{W}_{l, l+1} \right) \mathbf{D}_{N}^- \left(\mathbf{u}_N^\text{trgt} - \mathbf{u}_N^- \right) \left(\mathbf{r}_{k-1}^-\right)^T + \mathcal{O}(\lambda^2),
\end{align}
where we used that, to first order, the output layer error factor is $\mathbf{e}_N^- = \lambda \, \mathbf{D}^-_N \left(\mathbf{u}_N^\text{trgt} - \mathbf{u}_N^- \right) + \mathcal{O}(\lambda^2)$. Hence, up to a factor of $\lambda^{N-k}$ which can be absorbed in the learning rate $\eta_{k,k-1}$, changes induced by synaptic plasticity are equal to the backprop learning rule \eqref{eq:Delta-W-bp} in the limit $\lambda \to 0$, provided that the top-down weights are set to the transpose of the corresponding feedforward weights, $\mathbf{W}_{k,k+1} = \mathbf{W}_{k+1,k}^T$. The `quasi-feedforward' condition $\lambda \to 0$ has also been invoked to relate backprop to two-phase contrastive Hebbian learning in Hopfield networks \citep{Xie2003}.

\vspace{1em}
\textbf{Interneuron plasticity.}
The analyses of the previous sections relied on the assumption that the synaptic weights to and from interneurons were set to their ideal values, cf.~Eqs.~\ref{eq:ideal-wpi} and \ref{eq:ideal-wip}. We now study the plasticity of the lateral microcircuit synapses and show that, under mild conditions, learning rules \eqref{eq:dW-IP} and \eqref{eq:dW-PI} yield the desired synaptic weight matrices.

We first study the learning of pyramidal-to-interneuron synapses $\mathbf{W}^{\intn\pn}_{k,k}$. To quantify the degree to which the weights deviate from their optimal setting, we introduce the convex loss function
\begin{equation}
  \mathcal{L}_k^{\intn\pn} = \frac{1}{2} \Tr \left\{ (\mathbf{W}^{\intn\pn*}_{k,k} - \mathbf{W}^{\intn\pn}_{k,k} )^T (\mathbf{W}^{\intn\pn*}_{k,k} - \mathbf{W}^{\intn\pn}_{k,k} ) \right\},
\end{equation}
where $\Tr(\mathbf{M})$ denotes the trace of matrix $\mathbf{M}$ and $\mathbf{W}^{\intn\pn*}_{k,k} = \frac{g_\bas + g_\leak}{g_\bas + g_\apic + g_\leak} \mathbf{W}_{k+1, k}^{\pn \pn}$, as defined in Eq.~\ref{eq:ideal-wip}.

Starting from the pyramidal-to-interneuron synaptic plasticity rule \eqref{eq:dW-IP}, we express the interneuron somatic potential in convex combination form \eqref{eq:intn-convcomb} and then expand to first order around $\hat{\mathbf{v}}_k^\intn$,
\begin{align}
  \Delta \mathbf{W}_{k,k}^{\intn\pn} &= \eta^{\intn\pn}_{k,k} \left(\phi(\mathbf{u}_k^\intn) - \phi(\hat{\mathbf{v}}_k^\intn) \right) \, (\mathbf{r}^\pn_k)^T \nonumber\\
  &= \eta^{\intn\pn}_{k,k} \, \lambda \, \mathbf{D}_k^{\intn\pn} \left(\mathbf{u}_{k+1}^\pn - \hat{\mathbf{v}}_k^\intn \right) (\mathbf{r}^\pn_k)^T + \mathcal{O}(\lambda^2) \nonumber\\
  &= \eta^{\intn\pn}_{k,k} \, \lambda \, \frac{g_\bas}{g_\leak + g_\bas} \, \mathbf{D}_k^{\intn\pn} \left(\mathbf{W}_{k,k}^{\intn\pn*} - \mathbf{W}_{k,k}^{\intn\pn} \right) \mathbf{Q}_k + \mathcal{O}(\lambda^2).
\end{align}
Matrix $\mathbf{Q}_k = \mathbf{r}_k^\pn \, (\mathbf{r}_k^\pn)^T$ denotes the outer product, and $\mathbf{D}_k^{\intn\pn}$ is a diagonal matrix with $i$-th diagonal entry equal to $\phi^\prime(\hat{\mathbf{v}}_{k,i}^{\intn})$.

For simplicity, we ignore fluctuations arising from the stochastic sequential presentation of patterns \citep{Bottou1998} and look only at the expected synaptic dynamics\footnote{This can be understood as a batch learning protocol, where weight changes are accumulated in the limit of many patterns before being effectively consolidated as a synaptic update.}. We absorb irrelevant scale factors and to avoid a vanishing update we rescale the learning rate $\hat{\eta}^{\intn\pn}_{k,k}$ by $\lambda^{-1}$. Then, taking the limit $\lambda \to 0$ as in the previous sections yields
\begin{align}
  \Expect \! \left[\Delta \mathbf{W}_{k,k}^{\intn\pn} \right] &= \hat{\eta}^{\intn\pn}_{k,k} \left(\mathbf{W}_{k,k}^{\intn\pn*} - \mathbf{W}_{k,k}^{\intn\pn} \right) \Expect \! \left[ \mathbf{D}_k^{\intn\pn} \mathbf{Q}_k \right]\nonumber\\
  &= - \hat{\eta}^{\intn\pn}_{k,k} \, \frac{\partial \mathcal{L}_k^{\intn\pn}}{\partial \mathbf{W}^{\intn\pn}_{k,k}} \, \Expect \! \left[ \mathbf{D}_k^{\intn\pn} \mathbf{Q}_k \right] \label{eq:expected-dW-IP}.
\end{align}
The expectation is taken over the pattern ensemble. In the last equality above, we used the fact that the gradient of $\mathcal{L}_k^{\intn\pn}$ with respect to the lateral weights $\mathbf{W}^{\intn\pn}_{k,k}$ is given by the difference $\mathbf{W}^{\intn\pn*}_{k,k} - \mathbf{W}^{\intn\pn}_{k,k}$.

As long as the expectation on the right-hand side of \eqref{eq:expected-dW-IP} is positive definite, the synaptic dynamics is within $90º$ of the gradient and thus leads to the unique minimum of $\mathcal{L}_k^{\intn\pn}$. This condition is easily met in practice. For linear neurons, it amounts to requiring that the correlation matrix $\Expect [\mathbf{Q}]$ is positive definite. In other words, the patterns have to span $\mathbb{R}^{N_k}$, with $N_k$ being the number of pyramidal neurons at layer $k$. This is fulfilled when uncorrelated background noise currents are present, and it is likely the case for deterministic networks solving nontrivial tasks. For nonlinear neurons with a saturating transfer function $\phi$, saturation can lead to a matrix numerically close to singular and slow down learning. A weight matrix initialization that sets the neurons operating far from saturation is therefore an appropriate choice.

A mathematical analysis of the coupled system defined by the various plasticity rules acting in concert is rather involved and beyond the scope of this note. However, the learning of apical-targetting interneuron-to-pyramidal synapses can be studied in isolation by invoking a separation of timescales argument. To proceed, we assume that pyramidal-to-interneuron synapses are ideally set, $\mathbf{W}_{k,k}^{\intn\pn} = \mathbf{W}_{k,k}^{\intn\pn*}$. In practice, this translates to a choice of a small effective learning rate for apical-targetting weights $\eta_{k,k}^{\pn\intn}$. Note that, indirectly, this requirement also imposes a constraint on how fast top-down pyramidal-to-pyramidal weights $\mathbf{W}_{k,k+1}^{\pn\pn}$ can evolve. This is the parameter regime explored in the main text simulations with plastic top-down weights.

We can then proceed in a similar manner to the previous analysis, as we briefly outline below. Recalling from Eq.~\ref{eq:ideal-wpi} that $\mathbf{W}^{\pn\intn*}_{k,k} = - \mathbf{W}^{\pn\pn}_{k,k+1}$, we define the loss function
\begin{equation}
  \mathcal{L}_k^{\pn\intn} = \frac{1}{2} \Tr \left\{ (\mathbf{W}^{\pn\intn*}_{k,k} - \mathbf{W}^{\pn\intn}_{k,k} )^T (\mathbf{W}^{\pn\intn*}_{k,k} - \mathbf{W}^{\pn\intn}_{k,k} ) \right\}.
\end{equation}
After some manipulation, as $\lambda \to 0$ the expected synaptic change can be written as
\begin{equation}
  \label{eq:expected-dW-PI} \Expect \! \left[\Delta \mathbf{W}_{k,k}^{\pn\intn} \right] = - \eta_{k,k}^{\pn\intn} \, \frac{\partial \mathcal{L}_k^{\pn\intn}}{\partial{\mathbf{W}_{k,k}^{\pn\intn}}} \,
  \Expect \! \left[ \mathbf{r}^\pn_{k+1} \, (\mathbf{r}_{k+1}^\pn)^T \right],
\end{equation}
which leads us to conclude that the weights converge to the appropriate values, provided that the correlation matrix of layer $k\!+\!1$ activity patterns is positive definite.

\end{document}